\documentclass[review]{elsarticle}

\usepackage{lineno}

\usepackage{textcomp} 
\usepackage{multirow} 
\usepackage{placeins} 
\usepackage{subfig} 
\usepackage[title]{appendix} 
\usepackage{amsmath, amsthm, amssymb, amsfonts}
\usepackage{verbatim}
\usepackage[usenames,dvipsnames]{color}

\modulolinenumbers[5]

\journal{Fluid Dynamics Research}

\begin{document}

\begin{frontmatter}

\title{Roughness effects in laminar pipe flow}

\author[add1]{Utku Senturk\corref{mycorrespondingauthor}}
\cortext[mycorrespondingauthor]{Corresponding author}
\ead{utku.senturk@ege.edu.tr}

\author[add2]{Alexander J. Smits}

\address[add1]{Department of Mechanical Engineering, Ege University, Izmir 35040, Turkey}
\address[add2]{Department of Mech. and Aerospace Eng., Princeton University, Princeton 08544, USA}

\begin{abstract}
The impact of wall roughness on fully developed laminar pipe flow is investigated numerically.  The roughness is comprised of square bars of varying size and pitch.  Results show that the inverse relation between the friction factor and the Reynolds number in smooth pipes still persists in rough pipes, regardless of the rib height and pitch. At a given Reynolds number, the friction factor varies quadratically with roughness height and linearly with roughness pitch. We propose a single correlation for the friction factor that successfully collapses the data.
\end{abstract}


\end{frontmatter}

\section{Introduction}

The established framework of friction factor analysis in fully developed pipe flow is due to the early work by Nikuradse \cite{nikuradse1933laws} which reported experiments on artificially roughened pipes.  Three distinct flow regimes were identified: a laminar flow regime where roughness has virtually no effect on the friction, and a transitional regime and a fully turbulent regime where roughness becomes a significant factor once it exceeds a critical size (\cite{tani1987turbulent,perry1969rough,shockling2006roughness,allen2007turbulent}).  However,  the effects of roughness in laminar flows has been the subject of great interest to the microscale flow community and it is now well established that roughness leads to significant departures from Nikuradse's initial proposal (\cite{kandlikar2005characterization, shen2006flow,gamrat2008experimental, gloss2010wall}).

For a fully developed flow in a circular pipe of diameter $D$, the friction factor is given by
\begin{equation}
f = \frac{4\tau _{w}}{\frac{1}{2}\rho V^{2}} = \frac{ \left( -{d p}/{dx} \right) D}{\frac{1}{2}\rho V^{2}}
\label{Darcy}
\end{equation}
where $\tau_w$ is the local wall shear stress, $\rho$ is the fluid density and $\nu$ is its kinematic viscosity, $dp/dx$ is the pressure gradient, and $V$ is the mean (bulk) velocity.  For smooth pipes,  $f=64/Re$ where the Reynolds number is defined as $Re=VD/\nu$. 

Kandlikar \textit{et al.} \cite{kandlikar2005characterization} suggested that the principal influence of roughness in channel flow was that the effective channel height became $d=D-2k$, where $k$ is the characteristic roughness height, and that their results could be correlated using a constricted friction factor defined by 
\begin{equation}
f_c=\frac{ \left( -{d p}/{dx} \right) d}{\frac{1}{2}\rho V'^{2}}
\label{fKand}
\end{equation}
where $V'=VD/d$ for a channel.  The definition of the Reynolds number for a channel remains unchanged since by mass conservation $VD=V'd$.  

Kandlikar \textit{et al.} also proposed that for pipe flows the same constricted friction factor should apply, but with 
$V'=VD^2/d^2$.  In addition, the effective (constricted) Reynolds number is now different from $Re$, and it is given by
\begin{equation}
Re_c= \frac{V'd}{\nu} = \frac{Re}{1-2k/D}
\label{ReRec}
\end{equation}
Hence, it was proposed that we can relate the rough pipe friction factor to the smooth pipe friction factor according to
\begin{equation}
\frac{f_c}{64/Re_c}=\frac{f}{64/Re} \left( 1-\frac{2k}{D} \right)^4
\label{fKandf}
\end{equation}
 
More recently, Liu \textit{et al.} \cite{liu_li_smits} investigated numerically a laminar channel flow  with square bar roughness of height $k$, width $w$ and pitch $\lambda$.   Here, $k/D$ was varied from 0.01 to 0.1, and $\lambda/w$ ranged from 2 to 8.  They found that, in addition to roughness height, pitch was also important in determining the friction factor, and that a good collapse of friction factors could be obtained by the correlation
\begin{equation}
f_{c}^{'}={f_c} \left( 1+c\frac{\lambda}{w} \right)
\label{liuScaling}
\end{equation}
with $c=0.0136$ suggested by the authors for the best fit.  At our suggestion, as based on our findings given below, they were able to show that $f_c$ was inversely proportional to $Re$, as is the case for smooth pipe and channel flow.  

In this paper, we present the results of a detailed numerical study on the influence of the relative height and the spacing of the roughness elements for a fully developed laminar pipe flow with regular, square roughness elements for various Reynolds numbers. Specifically, we look for a correlation  in the form
\begin{equation}
f=f(Re,k/D,\lambda/w)
\label{corrFunctional}
\end{equation}
which successfully captures such variations in $f$.

\section{Problem description}

\begin{figure}[t]
\captionsetup[subfloat]{farskip=1pt,captionskip=1pt}
\centering
\subfloat[]{%
  \includegraphics[width=0.6\linewidth]{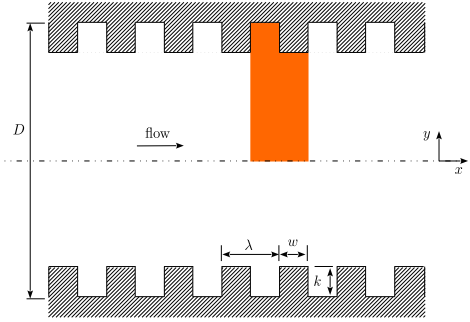}%
  }
\quad
\subfloat[]{%
  \includegraphics[width=0.6\linewidth]{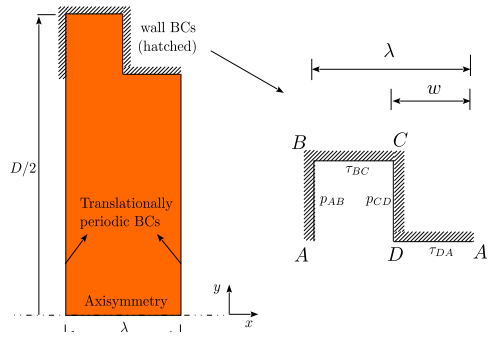}%
  }  
  \caption{Schematics of the problem: (a) Pipe section with square roughness elements. Colored portion is adequate for the translational periodic flow. (b) Simulation domain, assigned boundary conditions and pieces of the wall that contribute to pressure and shear.}
  \label{roughness}
\end{figure}

We consider steady, incompressible and fully developed flow in a horizontal pipe with square roughness elements on the wall (Figure \ref{roughness}a).  With these assumptions, we make use of the translational periodicity as described by Patankar \textit{et al.} \cite{patankar1977fully} and employed by Herwig  \textit{et al.}  \cite{herwig2008new}. For a brief description of the modeling with translational periodicity, see Appendix \ref{appendix}.

\begin{figure}
\centering
  \includegraphics[width=0.5\linewidth]{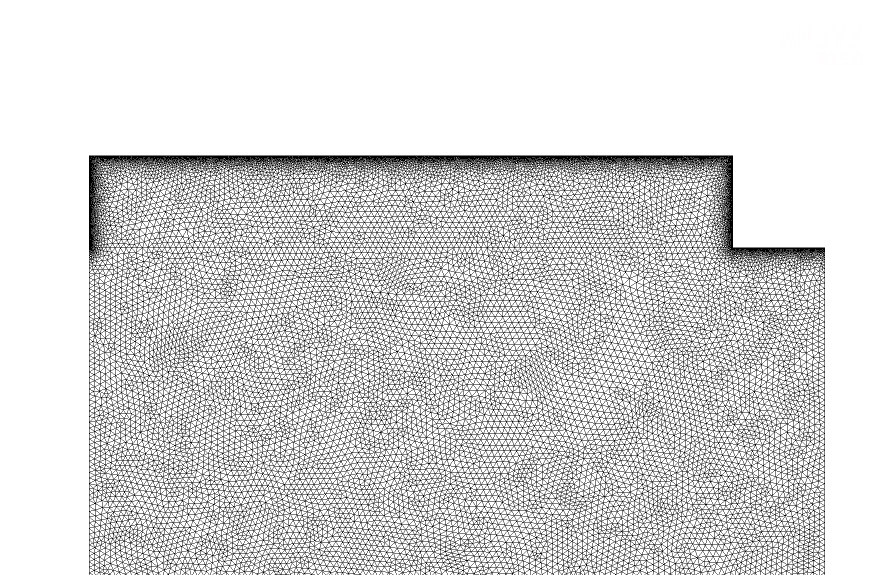}%
  \caption{An example of the unstructured finite volume grid used in simulations ($\lambda/w=8$ and $k/D=0.005$).}
  \label{grid}
\end{figure}

 The effect of the roughness height is investigated by varying $k/D$ from 0.005, to 0.01, 0.025 and 0.05. For each of these heights, four different roughness spacings are considered by setting $\lambda/w=2$, 4, 6 and 8 so that the smallest and largest spacings correspond to $d$-type and $k$-type roughness, respectively \citep{perry1969rough}. These simulations are repeated for 10 Reynolds numbers ranging from 200 to 2000. To analyze shear and pressure forces separately, the horizontal and vertical faces are considered separately, as shown in Figure \ref{roughness}b. Skin friction coefficients on horizontal face pieces ($BC$ and $DA$) are computed by,
\begin{equation}
C_{v}=\frac{\tau _{w}}{\frac{1}{2}\rho u_{m}^{2}}
\label{Cv}
\end{equation}
and the pressure coefficient that represents the net pressure force on the roughness element is computed using vertical face pieces ($AB$ and $CD$)
\begin{equation}
C_{p}=\frac{p _{AB}-p _{CD}}{\frac{1}{2}\rho u_{m}^{2}}
\label{Cp}
\end{equation}
Pressure and shear stress represent area-weighted average values in these definitions. Balance of forces on the domain gives,
\begin{equation}
(p_{up}-p_{down})A_{up}=(p_{AB}-p_{CD})A_{CD}+\tau_{BC}A_{BC}+\tau_{DA}A_{DA}
\label{forceBal}
\end{equation}
where we used $A_{up}=A_{down}$ and $A_{BC}=A_{CD}$ for the upstream wetted area and  roughness lateral area, respectively. Non-dimensionalizing with dynamic pressure and using the analogy with  Equation \ref{Darcy}, an effective friction factor can now be described:
\begin{equation}
f=\underbrace{\alpha C_p}_{f_p}+\underbrace{\gamma C_{v,BC}+\theta C_{v,DA}}_{f_v}
\label{fParts}
\end{equation}
Here, coefficients group several terms related to the geometry, for example $\alpha=DA_{CD}/(\lambda A_{up})$. An unstructured, non-uniform grid is used with triangular elements clustered near the channel walls ( Figure \ref{grid}). Using the pressure gradient obtained from a converged simulation, friction factor is found. The grid is successively refined until the change in the computed friction factor becomes negligibly small. Second-order accurate discretization schemes are used for the velocity (with upwinding) and the pressure. 

\section{Results}

\begin{figure}
\captionsetup[subfloat]{farskip=1pt,captionskip=1pt}
\centering
\subfloat[]{%
  \includegraphics[width=0.48\linewidth]{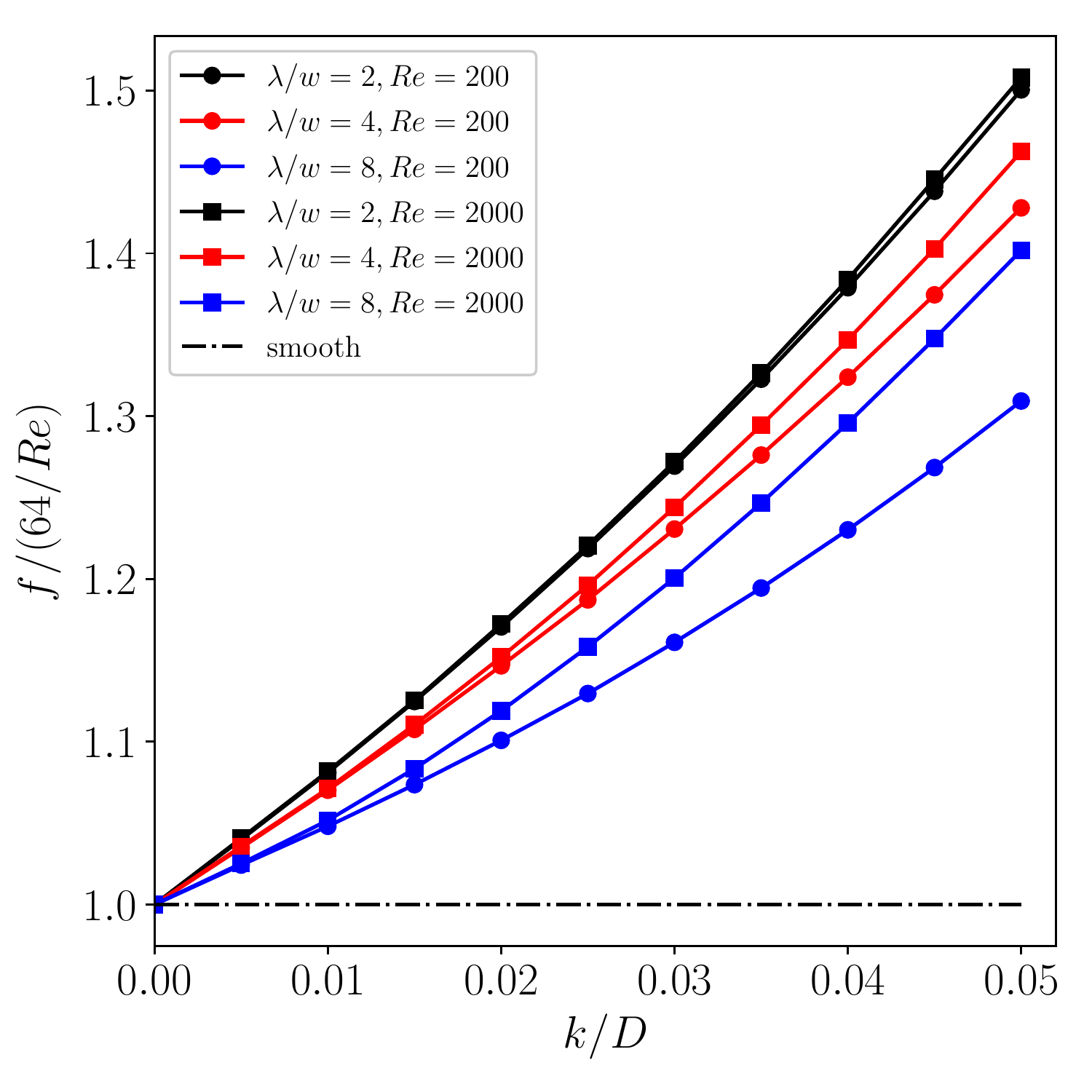}%
  }
  \subfloat[]{%
  \includegraphics[width=0.48\linewidth]{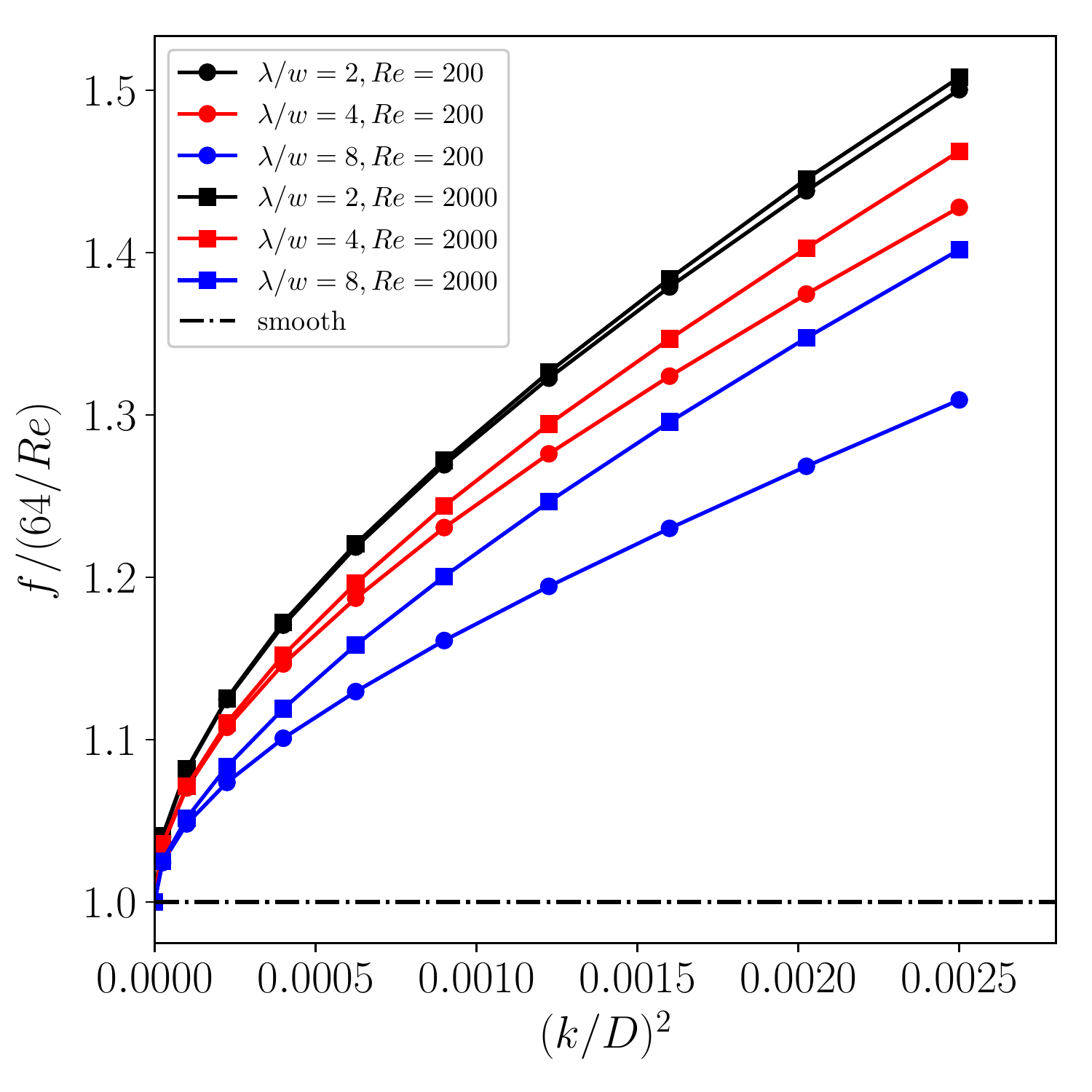}%
  }
  \caption{Computed friction factor normalized by that of a smooth pipe: (a) Friction factor increases quadratically with roughness height at any given spacing $\lambda/w$ and $Re$. (b) Results plotted as a function of $(k/D)^2$. }
  \label{fvskD}
\end{figure}

The effects of roughness height on the friction factor is given in Figure \ref{fvskD}. We show here only $Re=200$ and 2000 with $\lambda/w=2$, 4 and 8 for a clearer representation, although similar trends were found for all combinations of $Re$ and $\lambda/w$.  First remark in these curves is the relative increase in the friction factor compared to a smooth pipe ($64/Re$) as the roughness height increases which is in agreement with the previous findings (\cite{gamrat2008experimental, gloss2010wall, herwig2008new}). This relationship is quadratic and is found to be consistent in all of the spacings considered herein. Second, this effect amplifies as the spacing decreases, \textit{i.e.}, the roughness character migrates from a $k$-type and turns into a $d$-type. Third, there is an inverse relation between $Re$ and $f$ resembling smooth pipe flow: For a given roughness height and spacing, friction factor is larger at smaller $Re$. Finally, the effect of spacing is less at high $Re$. Therefore it can be concluded that introducing roughness on smooth pipe walls causes a significant increase in the effective friction due to the Equation \ref{fParts}, which depends quadratically on roughness height, $k/D$.

\begin{figure}
\captionsetup[subfloat]{farskip=1pt,captionskip=1pt}
\centering
\subfloat[]{%
  \includegraphics[width=0.48\linewidth]{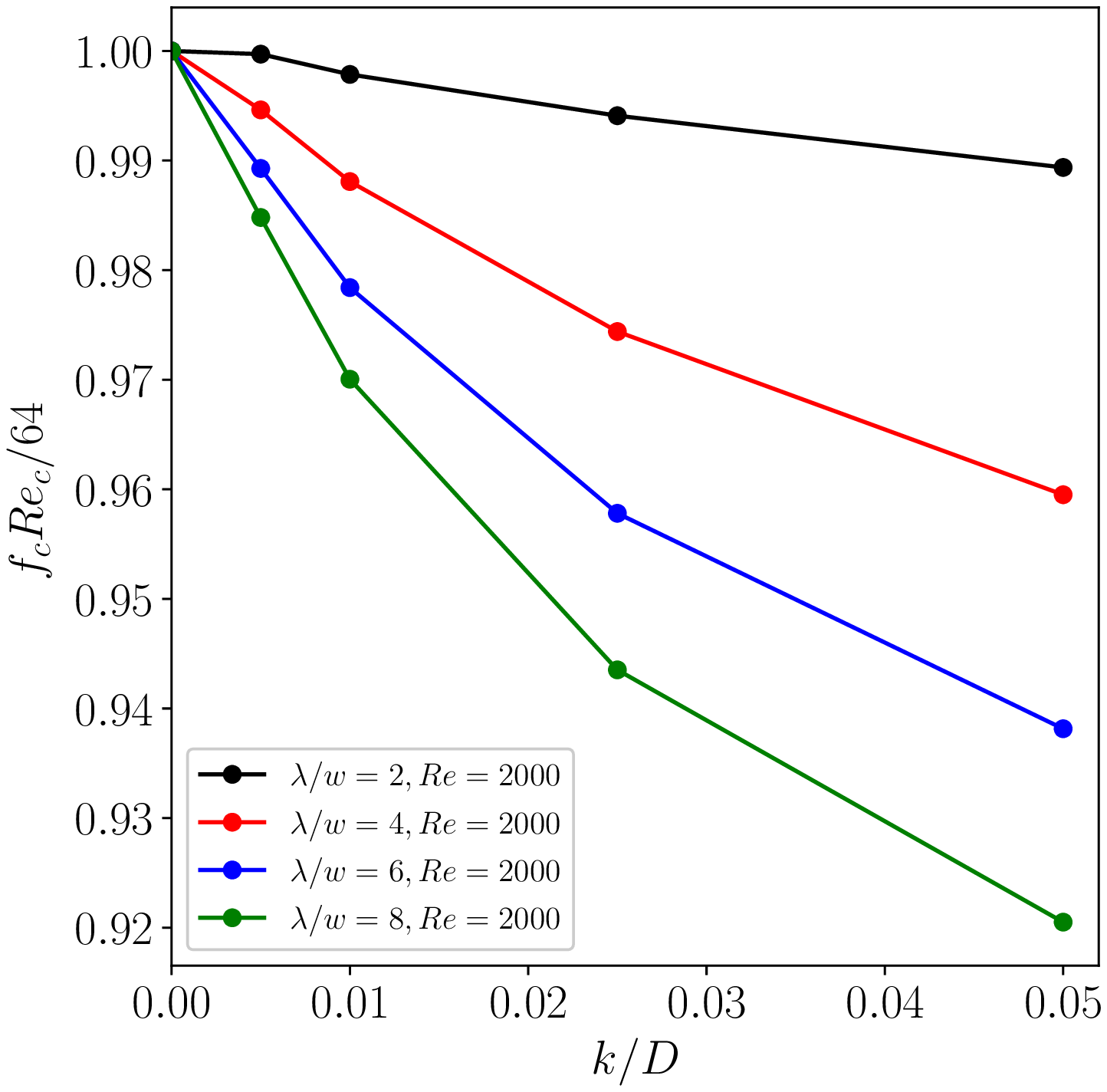}%
  }
  \subfloat[]{%
  \includegraphics[width=0.48\linewidth]{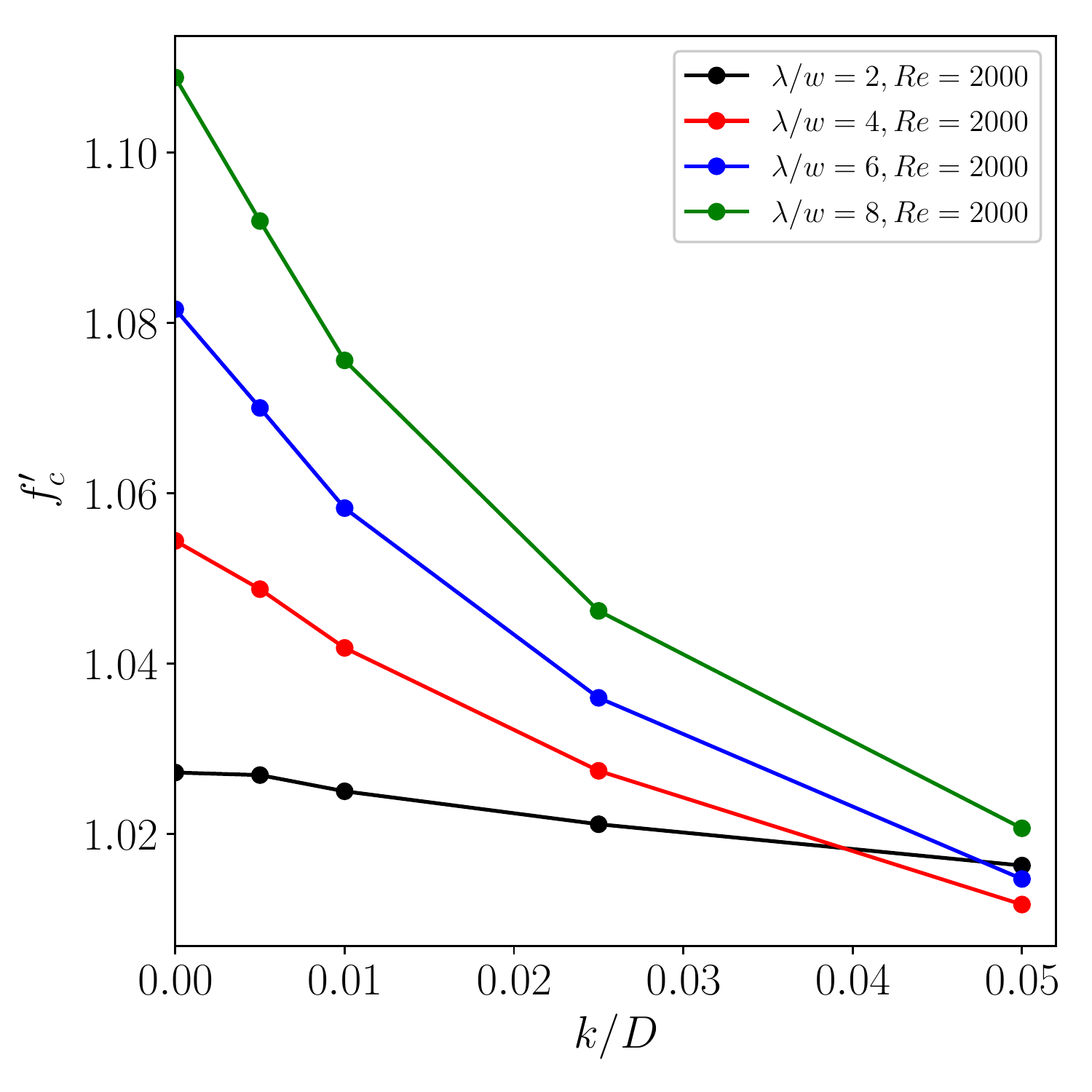}%
  }
  \caption{(a) Constricted friction factor as defined by Equation \ref{fKandf} following Kandlikar \textit{et al.} \cite{kandlikar2005characterization}. (b) Constricted friction factor as defined by Equation \ref{liuScaling} following Liu \textit{et al.} \cite{liu_li_smits}. }
  \label{Kand}
\end{figure}

We find that the data do not demonstrate the dependence on $k/D$ that follows from the concepts of constricted flow proposed by Kandlikar \textit{et al.} \cite{kandlikar2005characterization}, as expressed by Equation \ref{fKandf}).  Figure \ref{Kand} shows that there is a significant dependence on pitch, as found by Liu \textit{et al.} \cite{liu_li_smits}.  However, the linear scaling suggested by Liu \textit{et al.} for a channel, as given by Equation~\ref{liuScaling}, does not collapse our data for the pipe (see Figure~\ref{Kand}b.  Here, we used $c_1=0.0136$, as found by Liu \textit{et al.}, but it is obvious that no value for $c_1$ can be found to collapse the data properly.  

\begin{figure}
\captionsetup[subfloat]{farskip=1pt,captionskip=1pt}
\centering
\subfloat[]{%
  \includegraphics[width=0.48\linewidth]{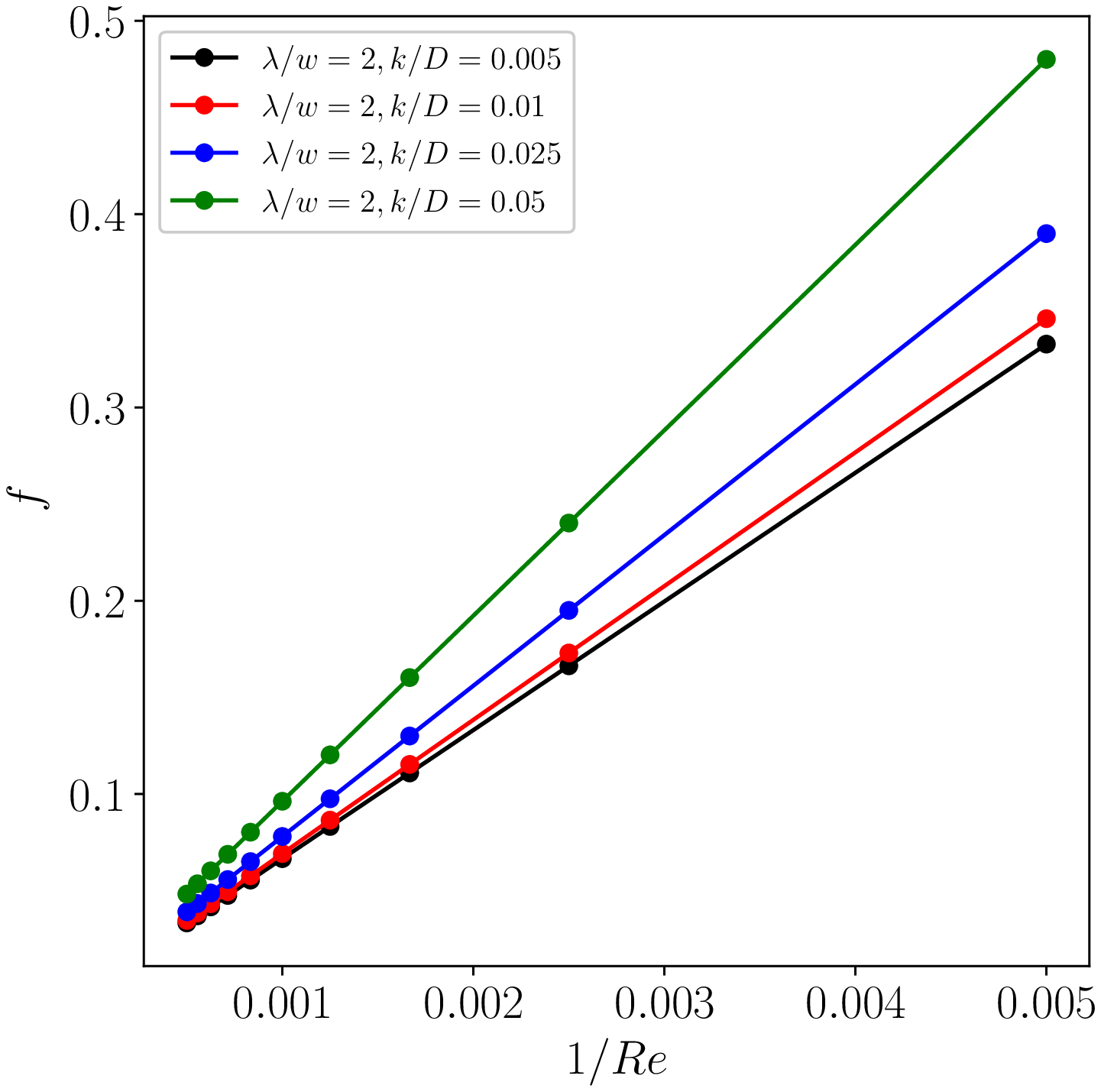}%
  }
  \subfloat[]{%
  \includegraphics[width=0.48\linewidth]{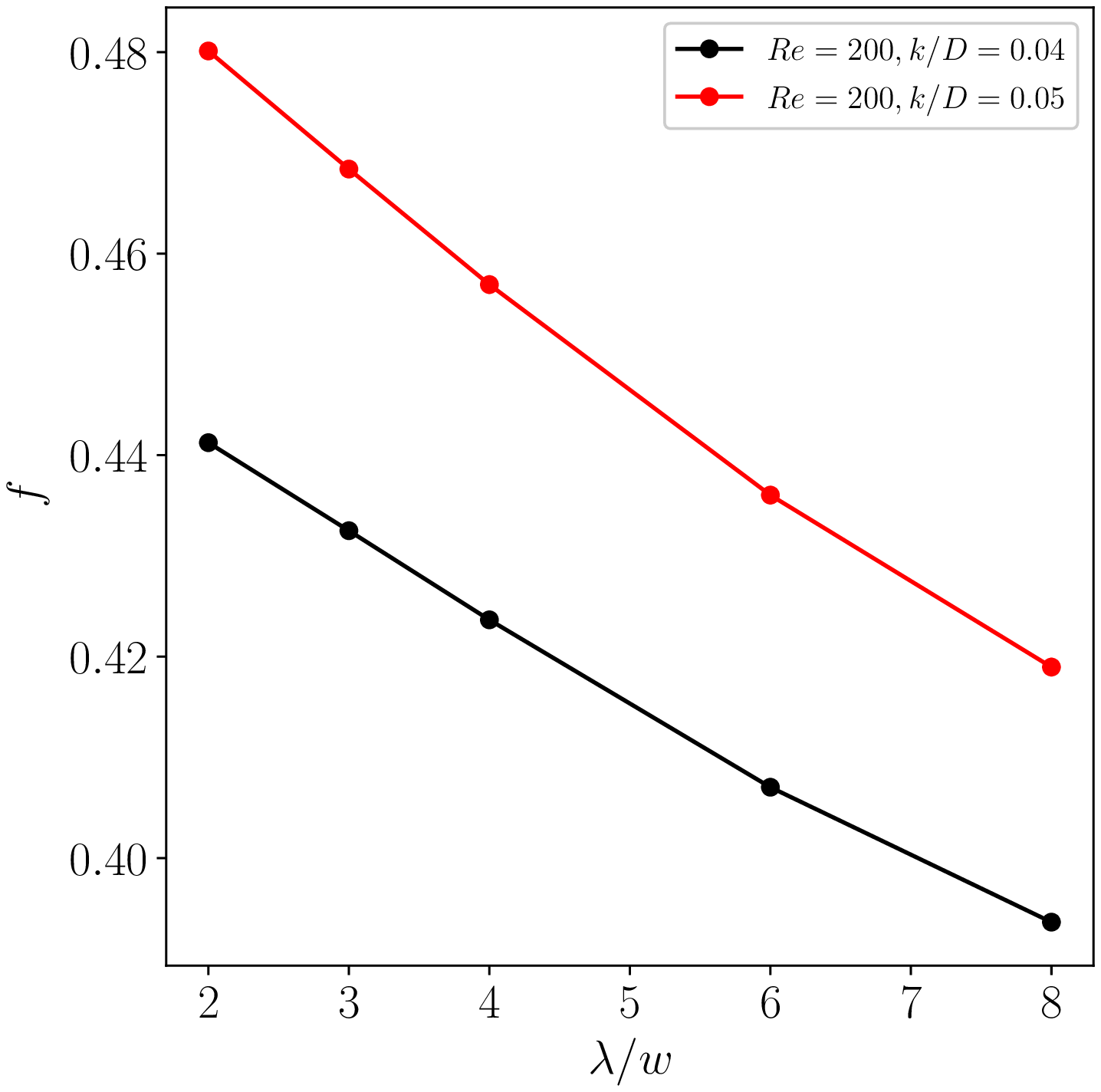}%
  }
  \caption{Dependence on $Re$ and $\lambda/w$: (a) Friction factor is inversely related to $Re$ as in smooth pipe flow. (b) As the spacing increases, friction factor decreases linearly.}
  \label{fvsRelw}
\end{figure}

Before addressing the influence of pitch on the friction factor in more detail, it is useful to consider first the effects of Reynolds number.  As illustrated in Figure \ref{fvsRelw}a for $\lambda/w=2$, we find that for any given $\lambda/w$ and $k/D$, the friction factor exhibits a linear relation with $1/Re$ precisely as in smooth pipe flow. At any given $Re$ and $\lambda/w$, friction factor increases with roughness height, as expected. 

\begin{figure}
\captionsetup[subfloat]{farskip=1pt,captionskip=1pt}
\centering
\subfloat[]{%
  \includegraphics[width=0.65\linewidth]{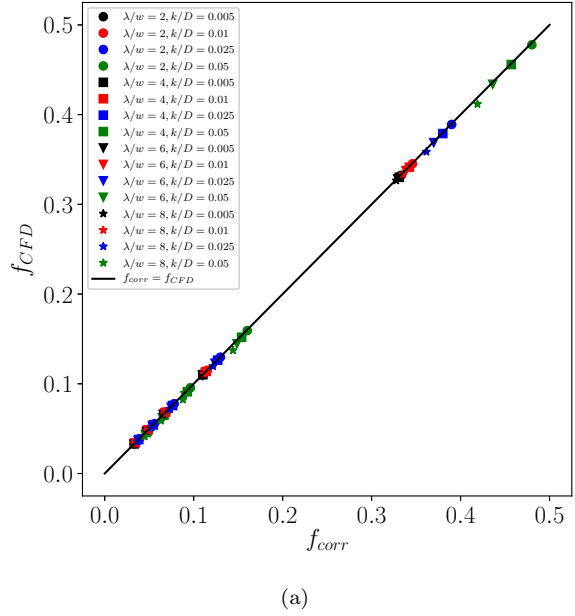}%
  }
  \caption{The correlation in Equation \ref{collapseCorr} proposed in this work successfully collapses all data.}
  \label{collapse}
\end{figure}

The dependence on pitch is shown in Figure \ref{fvsRelw}b for $Re=200$.  The trend where $f$ is a decreasing function of spacing with an almost linear trend is common across all Reynolds numbers.  In light of these findings, we use regression analysis to fit the data and propose the following correlation for $f$ on the parameters investigated here:
\begin{equation}
f=\frac{a_1\left ( \frac{k}{D} \right )^2+\left ( a_2+a_3 \frac{\lambda}{w} \right )\frac{k}{D}+64 }{Re}
\label{collapseCorr}
\end{equation}
Range of parameters in this work yielded $a_1=3000$, $a_2=520$ and  $a_3=-35$. Figure \ref{collapse} shows the excellent collapse obtained by the proposed correlation in Equation \ref{collapseCorr}. Largest error in this correlation is  7\% which is found at $k/D=0.05$ and $\lambda/w$=8.

\begin{figure}
\captionsetup[subfloat]{farskip=1pt,captionskip=1pt}
\centering
 \includegraphics[width=0.8\linewidth]{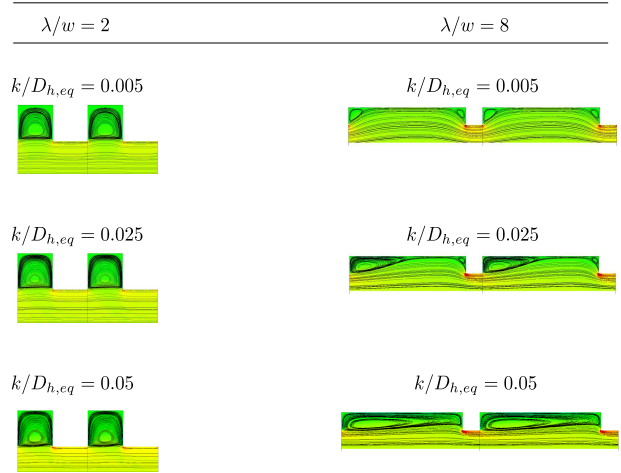}%
  \caption{Streamlines overlayed with the dimensionless vorticity contours ($\omega_{z}D_{h,eq}/u_{mean,eq}=[-0.02,0.02]$).}
  \label{streamline}
\end{figure}

\begin{figure}
\captionsetup[subfloat]{farskip=1pt,captionskip=1pt}
\centering
\subfloat[]{%
  \includegraphics[width=0.48\linewidth]{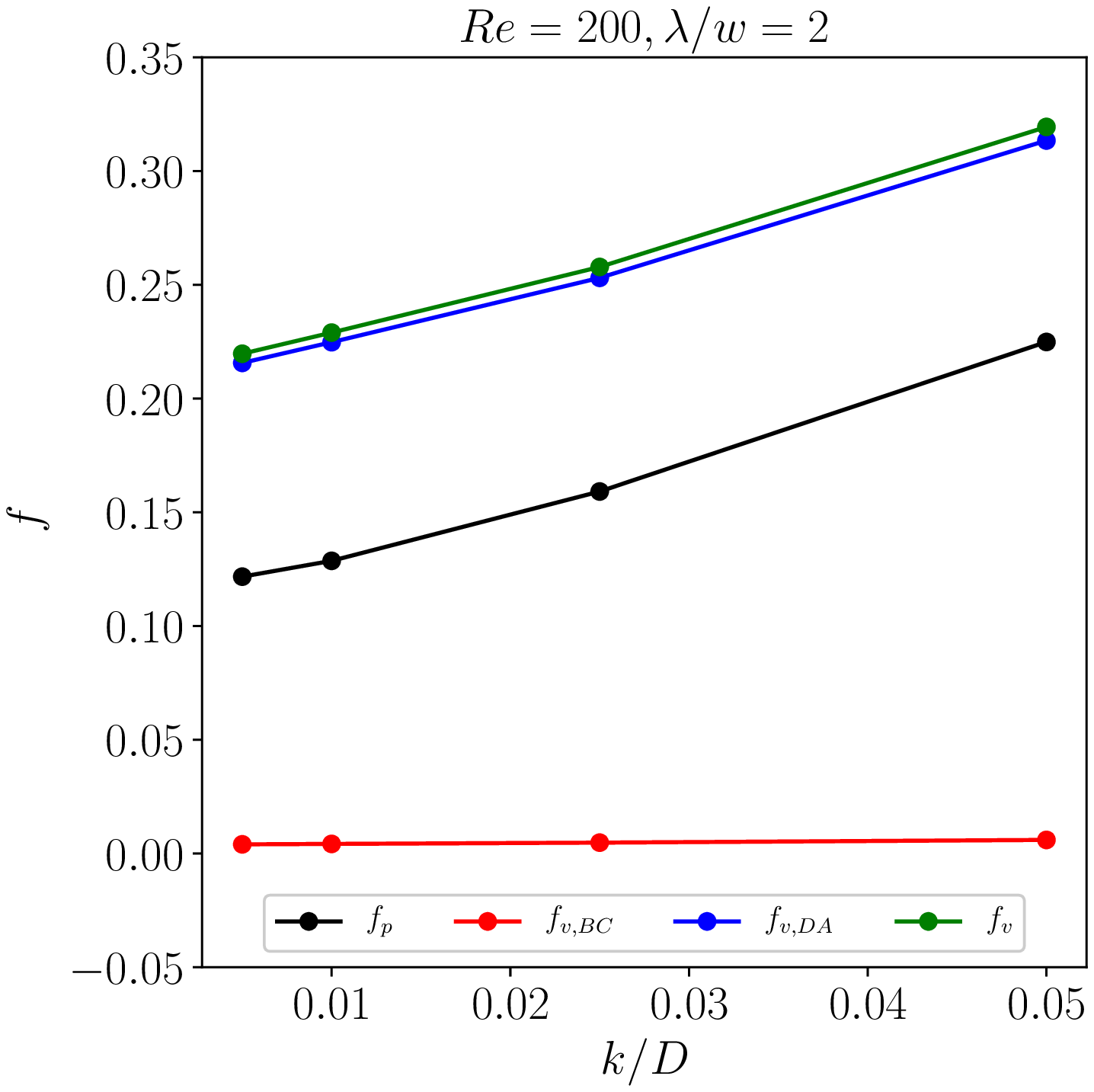}%
  }
  \subfloat[]{%
  \includegraphics[width=0.48\linewidth]{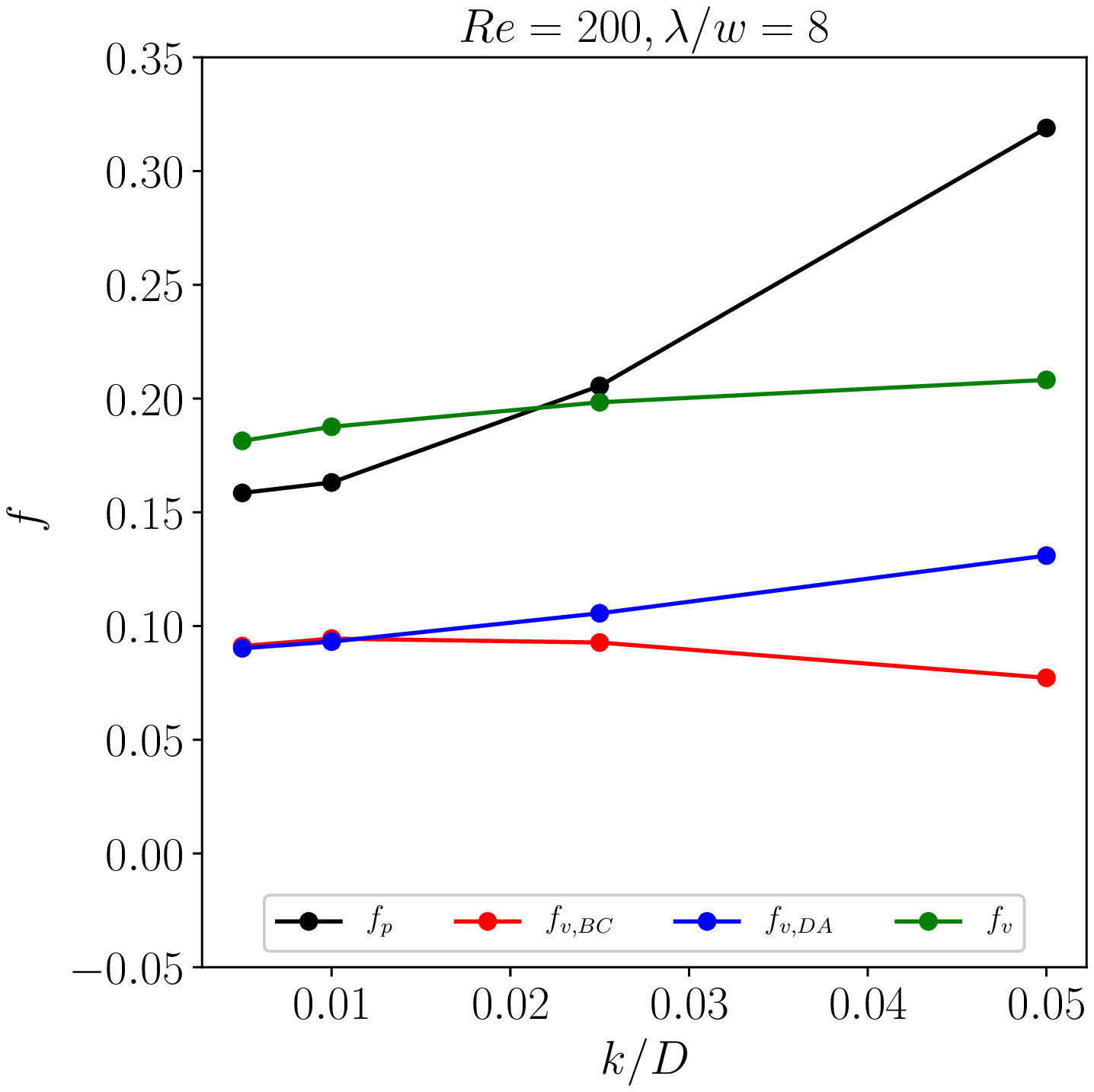}%
  }
  \caption{Pressure and shear contributions on $f$ found by using Equation \ref{fParts}: (a) $Re=200$ and $\lambda/w=2$. (b) $Re=200$ and $\lambda/w=8$.}
  \label{fPartsFig}
\end{figure}

Owing to the simplicity of the geometry, pressure and shear components of the friction factor can be analyzed using equation \ref{fParts}. Figure \ref{fPartsFig} shows that, for a $d$-type roughness element, shear on walls BC and DA contribute more than the net pressure on walls AB and CD for all roughness heights considered. This dependence is quadratic for all $f$ components. Moreover, shear on wall DA dominates the shear on wall BC, noting that both have equal areas for this spacing configuration.  However, for a $k$-type roughness element, we see a crossover at $k/D=0.025$ with pressure effects more pronounced at higher $k/D$. Shear on wall BC is now comparable to shear on DA as its area has increased in this configuration. Similar trends are observed at other Reynolds numbers. 

A deeper look in the flow field by means of vorticity contours and streamlines is given in Figure \ref{streamline} where two extremes of the roughness spacings are compared. In line with the turbulent flow roughness classification by Perry \textit{et al.} \cite{perry1969rough}, stable vortices seem to exist in the grooves of the $d$-type layout which exhibit qualitative similarity with the cavity flow. This behavior is consistent for all roughness heights considered. For the $k$-type layout, the roughness element behaves as a bluff body which creates a separation zone downstream, whose length increases with the relative roughness, in agreement with the turbulent flow simulations by Leonardi \textit{et al.} \cite{leonardi2007properties}. Reattachment of the flow on the channel walls is observed up to $k/D_{h,eq}= 0.025$. For $k/D_{h,eq}= 0.05$, the flow in the groove is completely separated. Additionally, contours of the dimensionless vorticity shows the high vorticity regions near the corners of the protruding edge of the roughness element in contrast to the relatively low vorticity regions in between the elements. In general, the vorticity in the vicinity of these corners becomes stronger as the relative height increases.

\section{Conclusion}
The effects of square roughness elements on laminar pipe flow has been presented. CFD simulations showed that Darcy friction factor increases quadratically with roughness height decreases linearly with pitch and inversely proportional to Reynolds number. A proper correlation was found that fits all data. 

\newpage

\begin{appendices}
\numberwithin{equation}{section}
\section{Modeling translational periodicity}\label{appendix}

The problem under consideration is the steady, incompressible, two dimensional and fully developed flow in a horizontal channel which contains square roughness elements on upper and lower walls (Figure \ref{roughness}a). With these assumptions, it is possible to make use of the translational periodicity as described by Patankar \textit{et al.} \cite{patankar1977fully} and employed by Herwig  \textit{et al.}  \cite{herwig2008new}. A brief description of the approach is as follows. Although the conditions related to the fully developed, smooth channel flow, $\partial u/\partial x=0$ and $v=0$, are no longer satisfied in a rough channel, it is still possible to define a fully developed regime in a periodic sense where the velocity field repeats itself as
\begin{equation}
\mathbf{V}(x,y)=\mathbf{V}(x+\lambda, y)=\mathbf{V}(x+2\lambda, y)=...
\label{RoughPerVel}
\end{equation}
where $\lambda$ is the periodic length in the direction of the flow. The pressure does not follow this behavior as it continuously decreases along the channel. However, the pressure drop does,
\begin{equation}
\Delta p=p(x,y)-p(x+\lambda ,y)=p(x+\lambda ,y)-p(x+2\lambda ,y)=...
\label{RoughPerDelP}
\end{equation}
and this is constant. Introducing the negative of the pressure gradient as
\begin{equation}
\beta =\frac{p(x,y)-p(x+\lambda ,y)}{\lambda}=-\frac{\partial p}{\partial x}
\label{beta}
\end{equation}
the actual pressure field can be decomposed into two components:
\begin{equation}
p(x,y)=-\beta x+\tilde{p}(x,y).
\label{pDecomposed}
\end{equation}
Incorporating into the governing equations yields,
\begin{equation}
\nabla \cdot \mathbf{V}=0
\label{cont}
\end{equation}
\begin{equation}
\nabla\cdot (\rho \mathbf{VV})=\beta \mathbf{i}-\nabla \tilde{p}+\mu \nabla^{2}\mathbf{V}.
\label{NSBeta}
\end{equation}
Hence, $\tilde{p}$ resembles the actual pressure in the Navier-Stokes equation and the $\beta$ merely acts as a source term. Numerical solution to equations \ref{cont} and \ref{NSBeta} are found using a finite volume-based commercial code ANSYS Fluent with a modified SIMPLE (semi-implicit method for pressure-linked equations) algoritm. Modification is due to the additional unknown $\beta$ which is guessed and succesively corrected throughout the iterative solution until the desired target mass flow rate input is matched.

\end{appendices}

\bibliography{references}

\end{document}